\documentclass[11pt]{article}
\usepackage[all,poly,ps,color]{xy}
\usepackage[utf8]{inputenc}
\usepackage[T1]{fontenc}
\usepackage{url,epsfig,amssymb}
\usepackage{color}
\usepackage{url}
\usepackage{grffile}
\usepackage{geometry}
\usepackage{sectsty}
\usepackage[normalem]{ulem}
\usepackage{amsmath}
\usepackage{apacite}
\usepackage{natbib}
\sectionfont{\sffamily\bfseries\upshape\large}
\subsectionfont{\sffamily\bfseries\upshape\normalsize}
\subsubsectionfont{\sffamily\bfseries\upshape\normalsize}
\makeatletter

  \renewcommand\@seccntformat[1]{\csname the#1\endcsname.\quad}

\makeatletter
\def\@maketitle{%
  \begin{center}%
  \let \footnote \thanks
    {\large \@title \par}%
    {\normalsize
      \begin{tabular}[t]{c}%
        \@author
      \end{tabular}\par}%
    {\small \@date}%
  \end{center}%
}
\makeatother

\title{\bf The 2008 election:  A preregistered replication analysis\footnote{To appear in {\em Statistics and Public Policy}.  We thank Catalist, the National Science Foundation, and the Office of Naval Research for partial support of this work, Macartan Humphreys for helpful comments, and Richard McElreath, Rob Trangucci, Jonah Gabry, Bob Carpenter, Ben Goodrich, and Daniel Lee for preparing the R and Stan functions that allowed us to easily fit our models.  Further details on our replication procedures appear in our online appendix, {\tt https://github.com/rayleigh/election\_stan\_analysis/blob/master/appendix.pdf}, and all our code can be found at {\tt https://github.com/rayleigh/election\_stan\_analysis}}\vspace{\baselineskip}}

\author{Rayleigh Lei\footnote{rxl2102@columbia.edu} \and Andrew Gelman\footnote{gelman@stat.columbia.edu} \and Yair Ghitza\footnote{yghitza@gmail.com}}

\date{\vspace{.1in} 4 Jul 2016}

\begin{document}

\maketitle

\begin{abstract}
We present an increasingly stringent set of replications of \cite{GandG}, a multilevel regression and poststratification analysis of polls from the 2008 U.S. presidential election campaign, focusing on a set of plots showing the estimated Republican vote share for whites and for all voters, as a function of income level in each of the states.

We start with a nearly-exact duplication that uses the posted code and changes only the model-fitting algorithm; we then replicate using already-analyzed data from 2004; and finally we set up preregistered replications using two surveys from 2008 that we had not previously looked at.  We have already learned from our preliminary, non-preregistered replication, which has revealed a potential problem with the published analysis of \cite{GandG}; it appears that our model may not sufficiently account for nonsampling error, and that some of the patterns presented in that earlier paper may simply reflect noise.

In addition to the substantive interest in validating earlier findings about demographics, geography, and voting, the present project serves as a demonstration of preregistration in a setting where the subject matter is historical (and thus the replication data exist before the preregistration plan is written) and where the analysis is exploratory (and thus a replication cannot be simply deemed successful or unsuccessful based on the statistical significance of some particular comparison).
\end{abstract}

\section{Background}\label{background}
Replication is central to scientific objectivity and is increasingly recognized to be important in social science as well.  In social science there has also been a movement toward {\em preregistration}, the specification of protocols for data collection, data analysis, and data processing ahead of time (see, e.g., \cite{Gonzales}), as a way of eliminating selection bias which has called into question inference from individual studies and entire literatures.  

A much-discussed cautionary example of such bias from psychology is the study of ``embodied cognition,'' in particular a paper by \cite{Bargh}, which has been cited over 3600 times but has recently been called into question.  After a series of failed replications \citep{doyen,wagenmakers}, it seems possible that the empirical results of that entire subfield can entirely be explained by a series of optimistic researchers capitalizing on noise.   In a more humble example, \cite{Nosek} recounted the story of their ``50 shades of gray'' experiment in which they obtained a striking result relating political extremism to color perception, a correlation that seemed to be strongly backed up by statistically significant $p$-values but which yielded null results under a careful preregistered replication. 

It has been increasingly recognized in recent years that non-preregistered studies can have problems arising from ``researcher degrees of freedom'' \citep{simmons} or the ``garden of forking paths'' \citep{gelmanloken}.  When a study is conducted in an open-ended fashion, researchers have many degrees of freedom to decide what data to collect, what data to exclude from their analysis, and what comparisons to perform, while the data are still coming in.  It becomes easy to get statistically significant $p$-values even if underlying effects are zero (or, more realistically, in settings with low signal-to-noise ratio so that statistically significant findings are likely to be highly exaggerated and often in the wrong direction; \cite{gelmancarlin}. 
As with the use of a hold-out sample in machine learning, external replication offers the promise of reducing such biases.

In political science, the term ``replication'' has traditionally been applied to the simple act of reproducing a published result using the identical data and code as used in the original analysis.  Anyone who works with real data will realize that this exercise is valuable and can catch problems with sloppy data analysis (for example, the Excel error of \cite{Reinhart}, or the ``gremlins'' paper of \cite{Tol}, which required nearly as many corrections as the number of points in its dataset; see \cite{gremlins}).  Re-examination of raw data can also expose mistakes, such as the survey data of \cite{LaCour}; see \cite{fakeStudy}.

But procedural replication does not address researcher degrees of freedom or forking paths.  To address these concerns it is helpful to have true replication with new data and a preregistered data-processing and analysis plan. \cite{Humphreys} discusses how formal preregistration can work with laboratory or field experiments and this seems like a promising approach:  We will not want every analysis to be preregistered but it is useful as an option, especially in studies that attempt to replicate controversial previously published claims.

Preregistered replication is more challenging in observational settings.  For one thing, observational data can sometimes not be replicated at all.  We cannot, for example, replicate an international relations study on a new sample of wars or recessions.  The other challenge is that some datasets are so well understood that it would be meaningless to talk about a preregistered data collection and analysis protocol.  Consider, for example, the much analyzed and much debated time series of economic growth and the party of the president  \citep{Bartels, Campbell,Comiskey}, a problem that can never again be virgin territory for statistical analysis.

All this has led to the awkward situation that we applaud the calls for preregistration of others' work but have never conducted a preregistered replication of our own \citep{preregistration}.

Recently, however, we have come across an opportunity to perform a preregistered replication of our own work.  In \cite{GandG} we reported the results of a statistical analysis of poll data from the Pew Research Center in the lead-up to the 2008 U.S. presidential election. Ghitza and Gelman performed several analyses; in the present paper we replicate one of them, an estimate of John McCain's share of the two-party by income, ethnicity, and state, as summarized in Figure 2 of that earlier article, which displays raw data and estimated McCain support as a function of five income categories for white voters and all voters in each of the 50 states.

We perform four replications of this analysis. Because we have already performed the first two replications and will describe them below, they are {\em not} preregistered:
\begin{enumerate}
\item A nearly-exact duplication, using the same data and model, just changing the statistical analysis slightly by fitting a fully Bayesian analysis in Stan in place of the marginal maximum likelihood estimate presented in \cite{GandG}.
\item A replication of the fully Bayesian analysis on a slightly different problem, the 2004 presidential election, using the Annenberg pre-election poll from that year.
\end{enumerate}
The above replications help us build trust in our method and smoke out any problems before then setting up the protocol for our two preregistered replications:
\begin{enumerate}
\setcounter{enumi}{2}
\item A replication using the fully Bayesian analysis on the {\em telephone sample} from the 2008 Annenberg pre-election survey.
\item A replication using the fully Bayesian analysis on the {\em internet sample} from the 2008 Annenberg pre-election survey.
\end{enumerate}
In Sections 1--4 of this article we give the results from analyses 1 and 2 above and set up our preregistration plan for analyses 3 and 4.  We will time-stamp our article up to that point and post on the internet.  Section 5 reports the reports the results of replications 3 and 4 and compares them to the earlier findings obtained from the Pew survey.

Replications in psychology are often performed because of suspicion or controversy about published findings, and the goal of such replications is often to resolve the controversy in some way.  The present replication is different.  The findings of \cite{GandG} have not been controversial; indeed that paper is largely methodological with no particular headline findings to confirm or reject.  Rather, the role of the present paper is to demonstrate the challenges of preregistration in a setting in which data are observational and analysis is complex.  In this case, as we suspect in many statistically-intensive problems, the design of a replication requires a surprising amount of effort.  This suggests why such replications are not performed more often and, we hope, motivates us to a future in which statistical workflows are specified in a more replicable fashion.

That said, there are some substantive implications that we would like our replication to address.  From a practical standpoint, the message of \cite{GandG} is that researchers can use multilevel regression and poststratification to make inferences about small subgroups of the population, for example the voting patterns of whites at different income levels within a state.  Our substantive focus will be to examine the results on income and voting from that published paper and see how they replicate with new data.  Unlike many replications, we are not checking particular comparison or coefficient to see if it remains statistically significant.  We hope this work is helpful to researchers in demonstrating replication in a more diffuse setting which, we believe, is characteristic of much social science research.

\section{Duplications and replications using existing data}\label{rep.old}

\cite{GandG} analyzed voter turnout and vote choice using a set of multilevel models predicting individual survey responses given income (divided into 5 categories), age (4 categories), ethnicity (white, black, hispanic, and other), and 51 states (including the District of Columbia).  For each binary outcome $y$ (for example, vote intention for the Republican or Democratic candidate, excluding respondents who are undecided or express other preferences), a logistic regression is fit, $\mbox{Pr}(y_i=1)=\mbox{logit}^{-1}(X_i\beta)$, where $X$ includes indicators for the demographic and geographic variables listed above, along with certain interactions of these main effects.  The coefficients $\beta$ are given a hierarchical prior distribution, in which batches of coefficients (``random effects,'' also called ``varying intercepts and slopes''; \cite{gelmanhill}) are assigned normal distributions with variances that are estimated from the data.
The model was used to create a predicted probability for each of the $5\times 4\times 4 \times 51$ {\em poststratification cells} $j$ corresponding to combinations of the demographic and geographic factors in the model.  Cells were then combined using Census data for the number of people or voters in each category.  This postratification step was performed using simple weighted averaging, for any subset $S$ of the population, computing
$\hat{\theta}_j=\sum_{j\in S}N_j\hat{\theta_j}/\sum_{j\in S}N_j$, where $\hat{\theta_j}$ represents the fitted $\mbox{Pr}(y=1)$ for people in cell $j$, and $N_j$ is the Census estimate of the population in the cell.

We performed our first replication, using the same data and altering the code only to fit a fully Bayesian version of our model in Stan \citep{stanRef}.  Compared to the earlier-fit model which performed marginal maximum likelihood, the only difference we noticed was that when the earlier point estimate calculated zero for a variable's terms, the Stan-calculated terms were nonzero. 
The practical differences between the two estimates were tiny, though, because when the marginal maximum likelihood estimate was zero for hierarchical variance parameters, the full Bayes estimates were small as well, and the differences were essentially nil when it came to predicted of turnout and vote choice proportions.

\begin{figure}
\centering
\includegraphics[width = \textwidth]{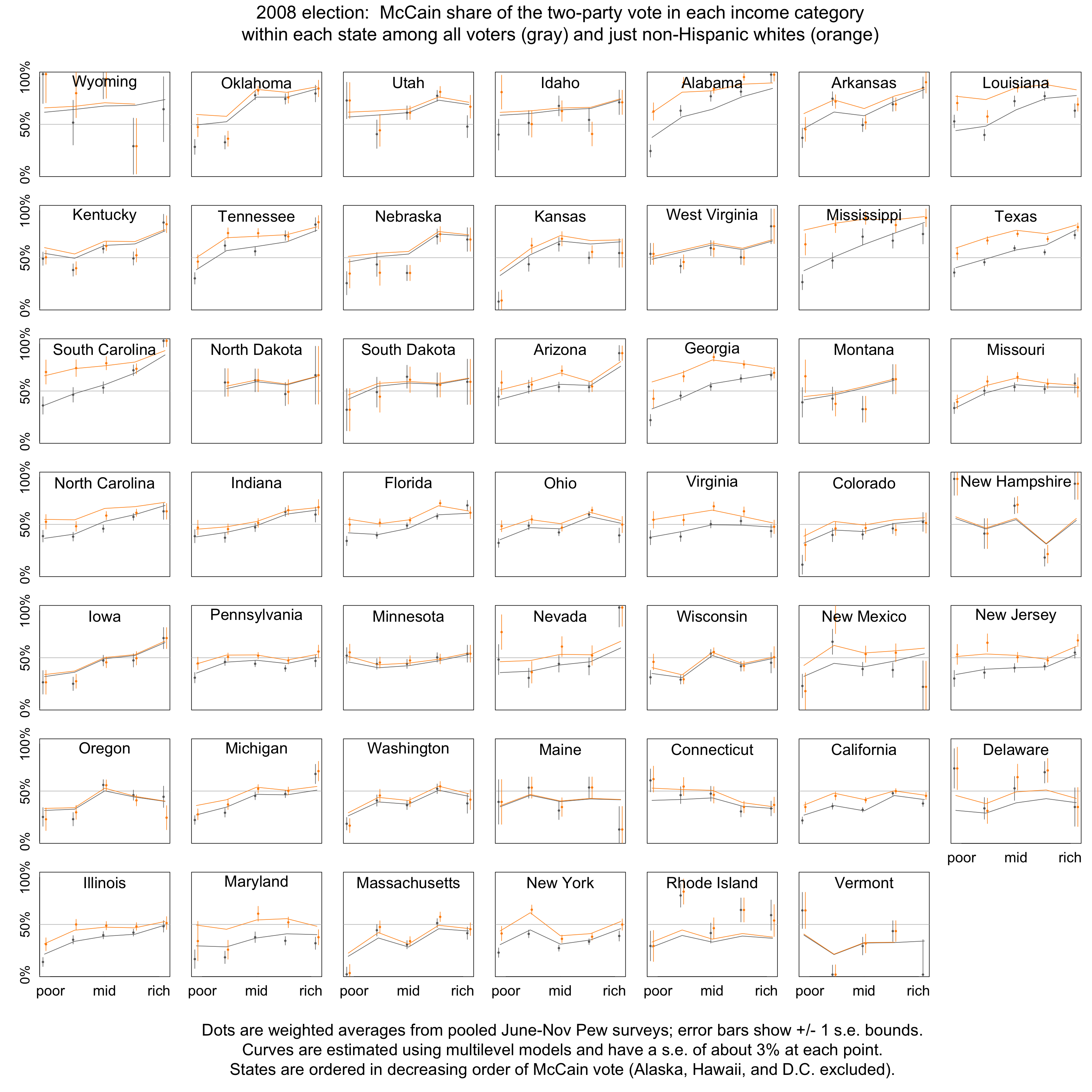}
\caption{\em Replication using Stan to fit the multilevel model to the 2008 Pew data, showing estimated McCain share of the two-party vote by state and income category for all voters (black) and just white voters (brown).   Changing to fully Bayesian analysis had little effect on the inferences, as can be seen by comparing to Figure 2 from \cite{GandG}.}
\label{rep1}
\end{figure}

Figure \ref{rep1} shows the results from the replications on the 2008 Pew data, with the only change being the switch to fully Bayesian inference.  The result is essentially the same as Figure 2 of \cite{GandG}, which is no surprise given the large sample size and large number of groups in the multilevel model.

\begin{figure}
\centering
\includegraphics[width = \textwidth]{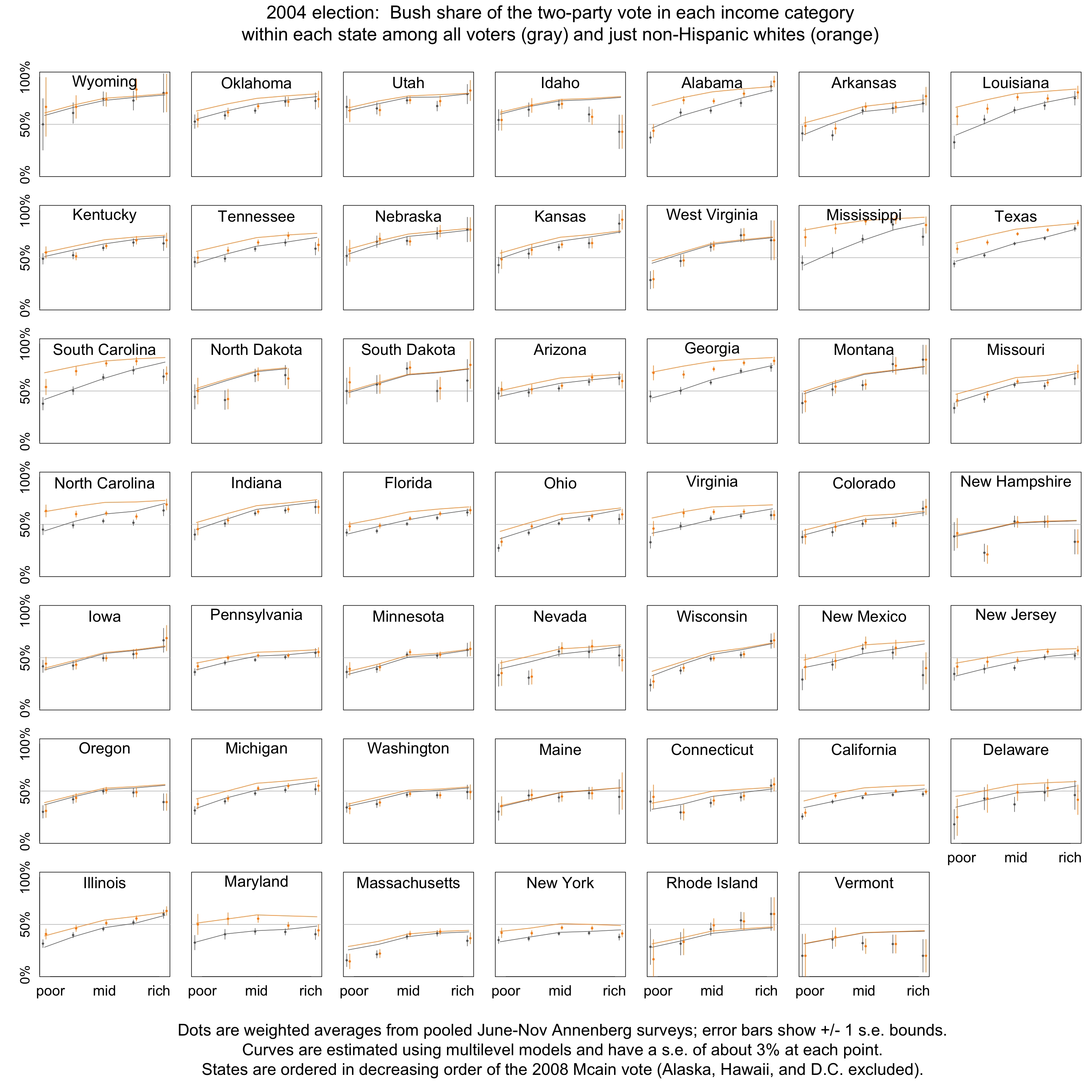}
\caption{\em Replication of Stan analysis using 2004 Annenberg survey, showing Bayesian estimates of George W. Bush's share of the two-party vote.  Compare to Figure \ref{rep1}, which shows the corresponding estimated Republican votes for 2008.}
\label{rep2}
\end{figure}

For our second replication, we applied our newly cleaned code to the 2004 Annenberg pre-election poll and produced Figure \ref{rep2}, which displays the raw data and Bayesian estimates for George W. Bush's vote share, by income, ethnicity, and state.

These results look reasonable, but we were struck by some differences as compared to the Pew 2008 analyses shown in Figure \ref{rep1}.  It should be no surprise to see changes in individual states, as the two elections were different, most notably among African Americans throughout the country and white voters in the South.  But we also notice a systematic difference:  the lines in Figure \ref{rep2} are much smoother than those in Figure \ref{rep1}. Our analysis from Annenberg 2004 shows a much more regular and monotonic pattern of income and voting by state, compared to our analysis from Pew 2008.  It's {\em possible} that this represents a real change but we think it is more likely a statistical artifact.

But what sort of artifact?  The sample size from Annenberg 2004 is 43,970, whereas Pew 2008 is based on only 19,170 respondents.  Based on this difference in sample size, we'd expect the Pew 2008 analysis to yield the smoother graphs: with its smaller sample, we would expect more pooling toward the logistic regression model, hence less jumpy curves.  Actually, though, as we see in Figures \ref{rep1} and \ref{rep2}, the curves from Pew (smaller sample) are jumpier than those from Annenberg (larger sample).

What is going on?  Again, it's possible that income was a less consistent predictor  of vote in 2008 than in 2004, and one could come up with explanations for specific patterns.  For example, consider the bump up in McCain support among whites in the second-lowest income category in Massachusetts and New York (see the bottom row of Figure \ref{rep2}).  Perhaps this can be understood as a disinclination of some voters in this group toward voting for an African American.  But it also seems plausible that many of the discordant patterns in Figure 2 simply represent noise---nonsampling error---that has not been accounted for the model.  The Pew and Annenberg surveys were conducted in different ways, and it is conceivable that Annenberg, which was entirely focused on the election campaign, could have more consistent responses during the months of data collection and a more representative sample than the Pew polls, which were designed with multiple purposes.

In any case, this discrepancy gives us another reason to perform a preregistered replication.  We will be able to compare different survey organizations using data from the same time period.

\section{Preregistering the new replications}

In preparation for the preregistered replications, we prepared the following files:
\begin{enumerate}
\item An R script to process the new data from the 2008 Annenberg surveys and also to load in the other information---Census summaries and state-by-state election results---needed for the multilevel regression and poststratification;
\item A Stan program to fit the multilevel model;
\item An R script to run the Stan program;
\item An R script to produce the equivalent of Figure \ref{rep1}.
\end{enumerate}
We plan to time stamp the present article and then perform the planned replications.
Having done so, we will compare the resulting graphs to Figure \ref{rep1}, which shows the estimates from the Pew data.   

\section{Discussion}

\subsection{Value of a preregistered replication plan}
In many of the most publicized examples of replication, the goal is to confirm or debunk some controversial existing research claim.  This present example is a bit different in that we began this study with no particular concern about the \cite{GandG} results, but a bit of replication would, we believe, give us a better sense of uncertainty about the details.  In addition, one can always be worried about opportunistic interpretations of statistical results.

The main point of the present paper, is to demonstrate that preregistered replication can be done, and it can be useful, even in a setting where the subject matter is historical (and thus the replication data exist before the preregistration plan is written) and where the analysis is exploratory (and thus a replication cannot be simply deemed successful or unsuccessful based on the statistical significance of some particular comparison). 

Laying out the details of this replication were instructive.  In practice it is not easy to replicate an existing analysis from several years ago.  Even in a case such as this where the data and code are accessible and so all the results can be reproduced, it can be a challenge to alter the analysis.  In this case a series of adaptations were required to move to a fully Bayesian analysis. In addition, it took some effort to prepare the Annenberg datasets for our new analysis.  And when we did the 2004 replication, we found interesting differences which suggested that there may indeed be problems with our published results (see Section \ref{rep.old}).

Our replication plan is preregistered and time-stamped but is not intended to be of forensic quality.  For example, there was nothing to stop us from secretly performing various analyses on the Annenberg data and using them to decide on the details of our purportedly preregistered design.  We did indeed have to crack open the new data in order to make coding decisions, and we can only offer our word that we did not make these choices based on outcomes. 

Our next step in this project is to analyze the Annenberg datasets from 2008 using our preregistered replication plan and compare to our earlier published graphs.  We plan to report these results in a follow-up paper.

\subsection{How to interpret the replication results?}

If all goes according to plan, performing the replications should take just a few minutes as they should merely involve running existing code on the two new, cleaned Annenberg 2008 pre-election polls, resulting in inferences and graphs for each dataset.  We will end up with two replications of Figure 2 from this paper (a figure which is itself essentially a duplication of Figure 2 from \cite{GandG}), showing data and estimated McCain vote share among whites and among all voters, for each of five income categories within each of the 48 contiguous states.

At that point, two challenges will arise.  The first is that, yes, the two new surveys represent independent data collections, but that all three surveys share biases:  they are all pre-election polls, and they all omit people who don’t respond to surveys.  Our replications will be valid {\em as replications} but it is possible that any patterns confirmed by the replications are still just artifacts driven by non-sampling error.  Thus we cannot automatically interpret patterns in our data, even if confirmed by replications, as representing truths about voters:  such a conclusion requires an extra-data assumption about representativeness of the surveys.

The other issue is that we have no predetermined criteria for ``success'' or ``failure'' of the replication.  Our plan is to remake and compare it to what we got from the earlier survey, and many comparisons can be made.  For example, voters in the second-lowest income category in several states (notably New York, Massachusetts, and California) appear to be noticeably more McCain-voting than those in the first and third income categories.  Is this something we should believe---it does, after all, appear to be reflected in the Bayesian estimates as well as in the raw data---or is it some artifact of the sample?  This is the kind of pattern we would like to check in a replication.  Confirmation of this pattern with the new surveys would not only increase our belief that this particular pattern from 2008 is real; it would also give us greater confidence in our inferential process.  Conversely, if the pattern does {\em not} replicate, we would be inclined to feel that Figure 2 overstates our certainty about the vote.

As the aforementioned example indicates, we plan our inspection of the replications to be somewhat open-ended, and we recognize that researcher degrees of freedom and forking paths will arise in our analysis.  Depending on what we see, we might well follow up with a larger Bayesian analysis including data from all three surveys, perhaps with survey-level error terms to capture differences between the samples beyond what could be explained by random sampling from a common population.  Another option would be to include the date of interview in the model, thus allowing for public-opinion shifts during the months in which these surveys were in the field.

\subsection{Presenting the replication results}
Once the present article (excluding Section \ref{results}) has been accepted for publication by {\em Statistics and Public Policy}, we will publicly post it online on the date shown on the first page.  We will then run the replcation, produce the two graphs as planned, and put those graphs and our discussion in Section \ref{results}, a section that will be blank in the initially posted version of the paper.  {\em Statistics and Public Policy} will then publish the entire article.

\section{Results from the preregistered analysis}\label{results}

[to be added upon acceptance and posting of Sections 1--4]

\begin{small}
\bibliographystyle{apacite}
\bibliography{refs.bib} 

\begin{thebibliography}{}

\bibitem [\protect \citeauthoryear {%
Bargh%
, Chen%
\BCBL {}\ \BBA {} Burrows%
}{%
Bargh%
\ \protect \BOthers {.}}{%
{\protect \APACyear {1996}}%
}]{%
Bargh}
\APACinsertmetastar {%
Bargh}%
\begin{APACrefauthors}%
Bargh, J\BPBI A.%
, Chen, M.%
\BCBL {}\ \BBA {} Burrows, L.%
\end{APACrefauthors}%
\unskip\
\newblock
\APACrefYearMonthDay{1996}{}{}.
\newblock
{\BBOQ}\APACrefatitle {Automaticity of social behavior: Direct effects of trait
  construct and stereotype-activation on action} {Automaticity of social
  behavior: Direct effects of trait construct and stereotype-activation on
  action}.{\BBCQ}
\newblock
\APACjournalVolNumPages{Journal of Personality and Social
  Psychology}{71}{}{230-244}.
\PrintBackRefs{\CurrentBib}

\bibitem [\protect \citeauthoryear {%
Bartels%
}{%
Bartels%
}{%
{\protect \APACyear {2008}}%
}]{%
Bartels}
\APACinsertmetastar {%
Bartels}%
\begin{APACrefauthors}%
Bartels, L\BPBI M.%
\end{APACrefauthors}%
\unskip\
\newblock
\APACrefYear{2008}.
\newblock
\APACrefbtitle {{Unequal} {Democracy}: {The} {Political} {Economy} {of} {the}
  {New} {Gilded} {Age}} {{Unequal} {Democracy}: {The} {Political} {Economy}
  {of} {the} {New} {Gilded} {Age}}.
\newblock
\APACaddressPublisher{}{Princeton University Press}.
\PrintBackRefs{\CurrentBib}

\bibitem [\protect \citeauthoryear {%
Campbell%
}{%
Campbell%
}{%
{\protect \APACyear {2011}}%
}]{%
Campbell}
\APACinsertmetastar {%
Campbell}%
\begin{APACrefauthors}%
Campbell, J\BPBI E.%
\end{APACrefauthors}%
\unskip\
\newblock
\APACrefYearMonthDay{2011}{}{}.
\newblock
{\BBOQ}\APACrefatitle {The Economic Records of the Presidents: Party
  Differences and Inherited Economic Conditions} {The economic records of the
  presidents: Party differences and inherited economic conditions}.{\BBCQ}
\newblock
\APACjournalVolNumPages{The Forum}{9}{}{1-29}.
\PrintBackRefs{\CurrentBib}

\bibitem [\protect \citeauthoryear {%
Comiskey%
\ \BBA {} Marsh%
}{%
Comiskey%
\ \BBA {} Marsh%
}{%
{\protect \APACyear {2012}}%
}]{%
Comiskey}
\APACinsertmetastar {%
Comiskey}%
\begin{APACrefauthors}%
Comiskey, M.%
\BCBT {}\ \BBA {} Marsh, L\BPBI C.%
\end{APACrefauthors}%
\unskip\
\newblock
\APACrefYearMonthDay{2012}{}{}.
\newblock
{\BBOQ}\APACrefatitle {Presidents, Parties, and the Business Cycle, 1949-2009}
  {Presidents, parties, and the business cycle, 1949-2009}.{\BBCQ}
\newblock
\APACjournalVolNumPages{Presidential Studies Quarterly}{42}{}{40-59}.
\PrintBackRefs{\CurrentBib}

\bibitem [\protect \citeauthoryear {%
Doyen%
, Klein%
, Pichon%
\BCBL {}\ \BBA {} Cleeremans%
}{%
Doyen%
\ \protect \BOthers {.}}{%
{\protect \APACyear {2012}}%
}]{%
doyen}
\APACinsertmetastar {%
doyen}%
\begin{APACrefauthors}%
Doyen, S.%
, Klein, O.%
, Pichon, C\BPBI L.%
\BCBL {}\ \BBA {} Cleeremans, A.%
\end{APACrefauthors}%
\unskip\
\newblock
\APACrefYearMonthDay{2012}{}{}.
\newblock
{\BBOQ}\APACrefatitle {Behavioral priming: It's all in the mind, but whose
  mind?} {Behavioral priming: It's all in the mind, but whose mind?}{\BBCQ}
\newblock
\APACjournalVolNumPages{PLoS ONE}{7}{}{e29081}.
\PrintBackRefs{\CurrentBib}

\bibitem [\protect \citeauthoryear {%
Gelman%
}{%
Gelman%
}{%
{\protect \APACyear {2013}}%
}]{%
preregistration}
\APACinsertmetastar {%
preregistration}%
\begin{APACrefauthors}%
Gelman, A.%
\end{APACrefauthors}%
\unskip\
\newblock
\APACrefYearMonthDay{2013}{}{}.
\newblock
{\BBOQ}\APACrefatitle {Preregistration of studies and mock reports}
  {Preregistration of studies and mock reports}.{\BBCQ}
\newblock
\APACjournalVolNumPages{Political Analysis}{21}{}{40-41}.
\PrintBackRefs{\CurrentBib}

\bibitem [\protect \citeauthoryear {%
Gelman%
}{%
Gelman%
}{%
{\protect \APACyear {2014}}%
}]{%
gremlins}
\APACinsertmetastar {%
gremlins}%
\begin{APACrefauthors}%
Gelman, A.%
\end{APACrefauthors}%
\unskip\
\newblock
\APACrefYearMonthDay{2014}{}{}.
\newblock
{\BBOQ}\APACrefatitle {A whole fleet of gremlins: Looking more carefully at
  {Richard} {Tol's} twice-corrected paper, ``{The} Economic Effects of Climate
  Change''} {A whole fleet of gremlins: Looking more carefully at {Richard}
  {Tol's} twice-corrected paper, ``{The} economic effects of climate
  change''}.{\BBCQ}
\newblock
\APAChowpublished {Statistical Modeling, Causal Inference, and Social Science
  blog, 27 May,
  \url{http://andrewgelman.com/2014/05/27/whole-fleet-gremlins-looking-carefully-richard-tols-twice-corrected-paper-economic-effects-climate-change/}}.
\PrintBackRefs{\CurrentBib}

\bibitem [\protect \citeauthoryear {%
Gelman%
}{%
Gelman%
}{%
{\protect \APACyear {2015}}%
}]{%
fakeStudy}
\APACinsertmetastar {%
fakeStudy}%
\begin{APACrefauthors}%
Gelman, A.%
\end{APACrefauthors}%
\unskip\
\newblock
\APACrefYearMonthDay{2015}{}{}.
\newblock
{\BBOQ}\APACrefatitle {Fake study on changing attitudes: Sometimes a claim that
  is too good to be true, isn't} {Fake study on changing attitudes: Sometimes a
  claim that is too good to be true, isn't}.{\BBCQ}
\newblock
\APAChowpublished {Monkey Cage blog, 20 May,
  \url{http://www.washingtonpost.com/blogs/monkey-cage/wp/2015/05/20/fake-study-on-changing-attitudes-sometimes-a-claim-that-is-too-good-to-be-true-isnt/}}.
\PrintBackRefs{\CurrentBib}

\bibitem [\protect \citeauthoryear {%
Gelman%
\ \BBA {} Carlin%
}{%
Gelman%
\ \BBA {} Carlin%
}{%
{\protect \APACyear {2014}}%
}]{%
gelmancarlin}
\APACinsertmetastar {%
gelmancarlin}%
\begin{APACrefauthors}%
Gelman, A.%
\BCBT {}\ \BBA {} Carlin, J.%
\end{APACrefauthors}%
\unskip\
\newblock
\APACrefYearMonthDay{2014}{}{}.
\newblock
{\BBOQ}\APACrefatitle {Beyond power calculations: Assessing Type S (sign) and
  Type M (magnitude) errors} {Beyond power calculations: Assessing type s
  (sign) and type m (magnitude) errors}.{\BBCQ}
\newblock
\APACjournalVolNumPages{Perspectives on Psychological Science}{9}{}{641-651}.
\PrintBackRefs{\CurrentBib}

\bibitem [\protect \citeauthoryear {%
Gelman%
\ \BBA {} Hill%
}{%
Gelman%
\ \BBA {} Hill%
}{%
{\protect \APACyear {2007}}%
}]{%
gelmanhill}
\APACinsertmetastar {%
gelmanhill}%
\begin{APACrefauthors}%
Gelman, A.%
\BCBT {}\ \BBA {} Hill, J.%
\end{APACrefauthors}%
\unskip\
\newblock
\APACrefYear{2007}.
\newblock
\APACrefbtitle {Data Analysis Using Regression and Multilevel/Hierarchical
  Models} {Data analysis using regression and multilevel/hierarchical models}.
\newblock
\APACaddressPublisher{}{Cambridge University Press}.
\PrintBackRefs{\CurrentBib}

\bibitem [\protect \citeauthoryear {%
Gelman%
\ \BBA {} Loken%
}{%
Gelman%
\ \BBA {} Loken%
}{%
{\protect \APACyear {2014}}%
}]{%
gelmanloken}
\APACinsertmetastar {%
gelmanloken}%
\begin{APACrefauthors}%
Gelman, A.%
\BCBT {}\ \BBA {} Loken, E.%
\end{APACrefauthors}%
\unskip\
\newblock
\APACrefYearMonthDay{2014}{}{}.
\newblock
{\BBOQ}\APACrefatitle {The statistical crisis in science} {The statistical
  crisis in science}.{\BBCQ}
\newblock
\APACjournalVolNumPages{American Scientist}{102}{}{460-465}.
\PrintBackRefs{\CurrentBib}

\bibitem [\protect \citeauthoryear {%
Ghitza%
\ \BBA {} Gelman%
}{%
Ghitza%
\ \BBA {} Gelman%
}{%
{\protect \APACyear {2013}}%
}]{%
GandG}
\APACinsertmetastar {%
GandG}%
\begin{APACrefauthors}%
Ghitza, Y.%
\BCBT {}\ \BBA {} Gelman, A.%
\end{APACrefauthors}%
\unskip\
\newblock
\APACrefYearMonthDay{2013}{}{}.
\newblock
{\BBOQ}\APACrefatitle {Deep Interactions with {MRP}: Election turnout and
  voting patterns among small electoral subgroups} {Deep interactions with
  {MRP}: Election turnout and voting patterns among small electoral
  subgroups}.{\BBCQ}
\newblock
\APACjournalVolNumPages{American Journal of Political Science}{57}{}{762-776}.
\PrintBackRefs{\CurrentBib}

\bibitem [\protect \citeauthoryear {%
Gonzales%
\ \BBA {} Cunningham%
}{%
Gonzales%
\ \BBA {} Cunningham%
}{%
{\protect \APACyear {2015}}%
}]{%
Gonzales}
\APACinsertmetastar {%
Gonzales}%
\begin{APACrefauthors}%
Gonzales, J\BPBI E.%
\BCBT {}\ \BBA {} Cunningham, C\BPBI A.%
\end{APACrefauthors}%
\unskip\
\newblock
\APACrefYearMonthDay{2015}{}{}.
\newblock
{\BBOQ}\APACrefatitle {The promise of pre-registration in psychological
  research} {The promise of pre-registration in psychological research}.{\BBCQ}
\newblock
\APACjournalVolNumPages{Psychological Science Agenda}{}{}{}.
\newblock
\begin{APACrefURL}
  \url{http://www.apa.org/science/about/psa/2015/08/pre-registration.aspx}
  \end{APACrefURL}
\PrintBackRefs{\CurrentBib}

\bibitem [\protect \citeauthoryear {%
Humphreys%
, de~la Sierra%
\BCBL {}\ \BBA {} van~der Windt%
}{%
Humphreys%
\ \protect \BOthers {.}}{%
{\protect \APACyear {2013}}%
}]{%
Humphreys}
\APACinsertmetastar {%
Humphreys}%
\begin{APACrefauthors}%
Humphreys, M.%
, de~la Sierra, R\BPBI S.%
\BCBL {}\ \BBA {} van~der Windt, P.%
\end{APACrefauthors}%
\unskip\
\newblock
\APACrefYearMonthDay{2013}{}{}.
\newblock
{\BBOQ}\APACrefatitle {Fishing, commitment, and communication: A proposal for
  comprehensive nonbinding research registration} {Fishing, commitment, and
  communication: A proposal for comprehensive nonbinding research
  registration}.{\BBCQ}
\newblock
\APACjournalVolNumPages{Political Analysis}{21}{}{1-20}.
\PrintBackRefs{\CurrentBib}

\bibitem [\protect \citeauthoryear {%
LaCour%
\ \BBA {} Green%
}{%
LaCour%
\ \BBA {} Green%
}{%
{\protect \APACyear {2014}}%
}]{%
LaCour}
\APACinsertmetastar {%
LaCour}%
\begin{APACrefauthors}%
LaCour, M\BPBI J.%
\BCBT {}\ \BBA {} Green, D\BPBI P.%
\end{APACrefauthors}%
\unskip\
\newblock
\APACrefYearMonthDay{2014}{}{}.
\newblock
{\BBOQ}\APACrefatitle {When contact changes minds: An experiment on
  transmission of support for gay equality} {When contact changes minds: An
  experiment on transmission of support for gay equality}.{\BBCQ}
\newblock
\APACjournalVolNumPages{Science}{346}{}{1366}.
\PrintBackRefs{\CurrentBib}

\bibitem [\protect \citeauthoryear {%
Nosek%
, Spies%
\BCBL {}\ \BBA {} Motyl%
}{%
Nosek%
\ \protect \BOthers {.}}{%
{\protect \APACyear {2012}}%
}]{%
Nosek}
\APACinsertmetastar {%
Nosek}%
\begin{APACrefauthors}%
Nosek, B\BPBI A.%
, Spies, J\BPBI R.%
\BCBL {}\ \BBA {} Motyl, M.%
\end{APACrefauthors}%
\unskip\
\newblock
\APACrefYearMonthDay{2012}{}{}.
\newblock
{\BBOQ}\APACrefatitle {Scientific Utopia {II}. {Restructuring} Incentives and
  Practices to Promote Truth Over Publishability} {Scientific utopia {II}.
  {Restructuring} incentives and practices to promote truth over
  publishability}.{\BBCQ}
\newblock
\APACjournalVolNumPages{Perspectives on Psychological Science}{7}{}{615-631}.
\PrintBackRefs{\CurrentBib}

\bibitem [\protect \citeauthoryear {%
Reinhart%
\ \BBA {} Rogoff%
}{%
Reinhart%
\ \BBA {} Rogoff%
}{%
{\protect \APACyear {2010}}%
}]{%
Reinhart}
\APACinsertmetastar {%
Reinhart}%
\begin{APACrefauthors}%
Reinhart, C\BPBI M.%
\BCBT {}\ \BBA {} Rogoff, K\BPBI S.%
\end{APACrefauthors}%
\unskip\
\newblock
\APACrefYearMonthDay{2010}{}{}.
\newblock
{\BBOQ}\APACrefatitle {Growth in a time of debt} {Growth in a time of
  debt}.{\BBCQ}
\newblock
\APACjournalVolNumPages{American Economic Review: Papers \&
  Proceedings}{100}{}{573-578}.
\PrintBackRefs{\CurrentBib}

\bibitem [\protect \citeauthoryear {%
Simmons%
, Nelson%
\BCBL {}\ \BBA {} Simonsohn%
}{%
Simmons%
\ \protect \BOthers {.}}{%
{\protect \APACyear {2011}}%
}]{%
simmons}
\APACinsertmetastar {%
simmons}%
\begin{APACrefauthors}%
Simmons, J.%
, Nelson, L.%
\BCBL {}\ \BBA {} Simonsohn, U.%
\end{APACrefauthors}%
\unskip\
\newblock
\APACrefYearMonthDay{2011}{}{}.
\newblock
{\BBOQ}\APACrefatitle {False-positive psychology: Undisclosed flexibility in
  data collection and analysis allow presenting anything as significant}
  {False-positive psychology: Undisclosed flexibility in data collection and
  analysis allow presenting anything as significant}.{\BBCQ}
\newblock
\APACjournalVolNumPages{Psychological Science}{22}{}{1359-1366}.
\PrintBackRefs{\CurrentBib}

\bibitem [\protect \citeauthoryear {%
{Stan Development Team}%
}{%
{Stan Development Team}%
}{%
{\protect \APACyear {2015}}%
}]{%
stanRef}
\APACinsertmetastar {%
stanRef}%
\begin{APACrefauthors}%
{Stan Development Team}.%
\end{APACrefauthors}%
\unskip\
\newblock
\APACrefYearMonthDay{2015}{}{}.
\newblock
{\BBOQ}\APACrefatitle {Stan Modeling Language: User's Guide and Reference
  Manual} {Stan modeling language: User's guide and reference manual}{\BBCQ}\
  (\PrintOrdinal{2.7}\ \BEd)\ [\bibcomputersoftwaremanual].
\PrintBackRefs{\CurrentBib}

\bibitem [\protect \citeauthoryear {%
Tol%
}{%
Tol%
}{%
{\protect \APACyear {2009}}%
}]{%
Tol}
\APACinsertmetastar {%
Tol}%
\begin{APACrefauthors}%
Tol, R.%
\end{APACrefauthors}%
\unskip\
\newblock
\APACrefYearMonthDay{2009}{}{}.
\newblock
{\BBOQ}\APACrefatitle {The economic effects of climate change} {The economic
  effects of climate change}.{\BBCQ}
\newblock
\APACjournalVolNumPages{Journal of Economic Perspectives}{23}{}{29-51}.
\PrintBackRefs{\CurrentBib}

\bibitem [\protect \citeauthoryear {%
Wagenmakers%
, Wetzels%
, Borsboom%
, Kievit%
\BCBL {}\ \BBA {} van~der Maas%
}{%
Wagenmakers%
\ \protect \BOthers {.}}{%
{\protect \APACyear {2015}}%
}]{%
wagenmakers}
\APACinsertmetastar {%
wagenmakers}%
\begin{APACrefauthors}%
Wagenmakers, E\BPBI J.%
, Wetzels, R.%
, Borsboom, D.%
, Kievit, R.%
\BCBL {}\ \BBA {} van~der Maas, H\BPBI L\BPBI J.%
\end{APACrefauthors}%
\unskip\
\newblock
\APACrefYearMonthDay{2015}{}{}.
\newblock
{\BBOQ}\APACrefatitle {A skeptical eye on psi} {A skeptical eye on psi}.{\BBCQ}
\newblock
\APACjournalVolNumPages{Extrasensory Perception: Support, Skepticism, and
  Science}{}{}{153-176}.
\PrintBackRefs{\CurrentBib}

\end{thebibliography}
\end{small}
\end{document}